\documentstyle[sprocl]{article}
\input psfig.tex    

\def\be{\begin{equation}}
\def\ee{\end{equation}}
\def\bea{\begin{eqnarray}}
\def\eea{\end{eqnarray}}

\begin{document}

\title{PHYSICS OF PHOTON DIFFUSION FOR ACTIVE SOURCES}

\author{R.A. Battye}

\address{Theoretical Physics Group, Blackett Laboratory, Imperial College \\
Prince Consort Road, London SW7 2BZ. U.K. }

\maketitle\abstracts{The physics of photon diffusion for the case of active sources is very different to that of passive sources. Fluctuations created just before the time of last scattering allow anisotropy to be created on scales much smaller than allowed by standard Silk damping. We develop a formalism to treat this effect and investigate its consequences using simple examples.}

Understanding the implications of active sources models, such as topological defects, on the cosmic microwave background is very exciting area of research, since accurate measurements of the CMB over a wide range of scales will soon be available, enabling us to constrain or rule out whole classes of theories, with profound implications for our understanding of physics at high energies. Here, we present a summary of a study of a particular aspect of this subject : the effects of photon diffusion or `Silk damping' \cite{silk}. We will extend previous work on passive \cite{HS} and active \cite{MACFa,MACFb} theories. A more thorough discussion of this subject is currently in press \cite{BAT}. 

The crucial period for understanding these effects is that just before last scattering. During this epoch the acoustic oscillations in the photon-baryon fluid are damped by the increasing mean free path of the photons. Active sources create fluctuations after the onset of this regime, which will receive less damping than those created before it. In particular, those created just before the time of last scattering will receive virtually no damping at all. If perturbations are created on all scales above the core size, as is thought to be the case for defect models, then it will be possible for anisotropy on small angular scales to remain to the present day. We will see that it is not sufficient to model these effects with a simple multiplication by an exponential suppression factor across all scales. Rather it requires careful consideration of the time at which fluctuations are created. Simple estimates will show that there are potentially important modifications to peak heights and also power-law suppression at small angular scales, rather than exponential.

In the region of angular scales of interest, the anisotropy can be estimated by investigating the behaviour of the intrinsic component of the anisotropy $\Theta_0$, since it is this which is imprinted on the microwave background at last scattering, leading to Doppler or Sakharov peaks. An equation for $\widehat\Theta_0=\Theta_0+\Phi$ which includes the effects of photon diffusion at first order is
\be
{\ddot{\widehat\Theta_0}}+\left({\dot R\over 1+R}+ {8\over 27}k^2{1\over\dot\kappa}{1\over 1+R}
\right){\dot{\widehat\Theta_0}} + {1\over 3}k^2{1\over 1+R}{\widehat\Theta_0} = H(\eta)={1\over 3}k^2\left({\Phi\over 1+R}-\Psi\right)\label{photon:eqn}
\ee
where $R=3\rho_{\rm b}/4\rho_{\gamma}$, $\dot\kappa$ is the differential optical depth due to Thomson scattering and $\Phi,\Psi$ are the gauge invariant gravitational potentials. Intuitively, one think of this as a forced-damped harmonic oscillator. Obviously, even in this very simple analogy no forcing effect can be damped until the it actually takes place.

One can solve (\ref{photon:eqn}) for large $k$ using the WKB approximation. If one also ignores the transient solution, which can be thought of as being due to passive fluctuations, then $\widehat\Theta_0$ is given by
\be
\left[1+R(\eta)\right]^{1/4}\widehat\Theta_0(\eta)={\sqrt{3}\over k}\int_{0}^{\eta} d\eta^{\prime} F(k,\eta,\eta^{\prime})
\ee
where 
\be
F(k,\eta,\eta^{\prime})=\left [1+R(\eta^{\prime})\right]^{3/4}e^{-k^2/k_{\rm s}^2(\eta,\eta^{\prime})} \sin\left[kr_{\rm s}(\eta)-kr_{\rm s}(\eta^{\prime})\right] H(\eta^{\prime})\,,
\ee
$r_s(\eta)$ is the sound horizon distance and $k_{\rm s}^{-1}(\eta_2,\eta_1)$ is the Silk damping length, below which photon diffusion removes anisotropy,
\be
r_{\rm s}(\eta)={1\over\sqrt{3}}\int_0^{\eta}{d\eta^{\prime}\over\sqrt{1+R(\eta^{\prime})}}\,,\quad k^{-2}_{\rm s}(\eta_2,\eta_1)={4\over 27}\int_{\eta_1}^{\eta_2}{d\eta^{\prime}\over\dot\kappa(\eta^{\prime})}{1\over 1+R(\eta^{\prime})}\,.
\ee
One can see immediately that the effect of photon diffusion is not an exponential suppression across all scales. Instead, the exponential suppression occurs inside the integral, with the damping scale being time dependent.

The effects of damping can be investigated by evaluating the power spectrum of of $\widehat\Theta_0$ using a numerical integration routine and the simple structure functions designed to represent the qualitative behaviour of, for example, strings or textures.
Here, we just present the results for a string structure function in figs.~1(a) and 1(b). Using a linear scale and a log-log scale illustrates the two effects of including this modified damping. One can see that the second, fourth and sixth peak heights are increased by 13\%, 25\% and 44\% respectively and the tail at large $x_*(=k\eta_*/\sqrt{3})$ is no longer exponential, rather it is power law. The observable consequence of this effect seems to be a modulation of peak heights and a power-law tail, very similar to the effects of baryon drag \cite{HSc}. In that case, the imperfect coupling between baryons and photons is responsible for an effective baseline shift. Here, what we are seeing is the sum of two components, one which is exponential, due to perturbations created before the onset of last scattering, and the other which is power law, due to perturbations created during the surface of last scattering.

\begin{figure}
\centerline{\psfig{figure=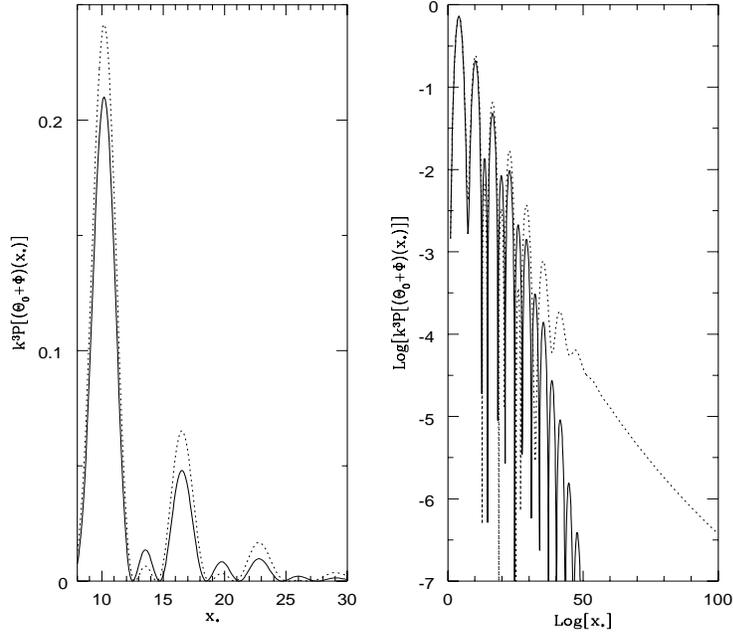,width=4.0in,height=3.6in}}
\caption{A comparison between the standard damping formalism (solid line) and the modified damping formalism (dotted line) for a string-like structure function : (a) on a linear scale and (b) on a log-log scale. Note that the linear scale starts at $x_*$=8 and hence ignores the first peak.}
\end{figure}

\end{document}